\documentclass[10pt,letterpaper,english]{article}
\usepackage[T1]{fontenc}
\usepackage[latin1]{inputenc}
\usepackage{amsmath}
\usepackage{graphicx}
\usepackage{amssymb}
\usepackage{array}
\makeatletter

\makeatletter
\usepackage{amscd}

\usepackage{amsfonts}

\usepackage{babel}\makeatother

\makeatother

\usepackage{babel}

\begin{document}
\newtheorem{proposition}{Proposition}[section] \newtheorem{definition}{Definition}[section]
\newtheorem{corollary}{Corollary}[section] \newtheorem{lemma}{Lemma}[section]
\newtheorem{theorem}{Theorem}[section] \newtheorem{example}{Example}[section]

\title{\textbf{An Application of Dirac's Interaction \\ Picture to Option Pricing.}}

\author{Mauricio Contreras G.\thanks{Universidad Metropolitana de Ciencias de la Educación UMCE, Chile, email: mauricio.contreras@umce.cl}}

\maketitle
\noindent
In this paper, the Dirac's quantum mechanical interaction picture is applied to option pricing to obtain a solution of the Black--Scholes equation in the presence of a time-dependent arbitrage bubble.
In particular, for the case of a call perturbed by a square bubble, an approximate solution (valid up third order in a perturbation series) is given in terms of the three first Greeks: Delta, Gamma, and Speed.
Then an exact solution is constructed in terms of all higher order  $S$-derivatives of the Black--Scholes formula.\\
It is also shown that the interacting Black--Scholes equation is invariant under a discrete transformation that interchanges the interest rate with the mean of the underlying asset and vice versa.
This implies that the interacting Black--Scholes equation can be written in a `low energy' and a `high energy' form, in such a way that the high-interaction limit of the low energy form corresponds to the weak-interaction limit of the high energy form.
One can apply a perturbative analysis to the high energy form to study the high-interaction limit of the low energy form.
\newpage
\section{Introduction}
\noindent
It is well known that the quantum mechanical evolution of a physical system can be found by using any one of three representations: the Schr\"{o}dinger picture, the Heisenberg picture, or the interaction picture \cite{Merzbacher, Sakurai}.
In particular, the Interaction or Dirac's picture is well suited to describe systems that have time-dependent forces.
This framework permits, for example, obtaining the solution for the wave-function, Green's functions, or scattering amplitudes of the correlated system as a perturbation series in the same free quantities.
The usual old fashioned form of this method that uses operator algebraic manipulations has been applied in the quantum field theoretic description of many-body systems \cite{Fetter, Abrikosov}, condensed matter \cite{Altland, Tsvelik}, and particle physics \cite{Roman, Itzykson}.
A fourth picture based on the Feynman's path integral formalism is capable of implementing the modern gauge theory model of the fundamental forces more easily.
This last scheme also permits developing a perturbation series, which is equivalent to that of Dirac's picture \cite{Bailin, Weinberg}.\\ \\
On the other hand, physicists have begun to apply their methods and mathematical machinery to disciplines outside of physics, such as finance and economics.
For example, one can find thermodynamic \cite{Karakatsanisa, Zhou} and statistical mechanical approaches \cite{Baaquie1, Baaquie2} applied to financial markets.
Also, different ideas have been proposed to understand the Black--Scholes model as a Quantum Mechanical one \cite{Haven}--\cite{contrerassemi}.
Quantum field theoretical methods have been applied to interest rate modeling \cite{BaaquieQFT1}--\cite{BaaquieQFT3} and even ideas from super-symmetry have been applied to option pricing \cite{susy1}--\cite{susy3}.
Even more, the theoretical optimal control description of economic systems can be understood as a constrained classical system in phase space \cite{contrerasDynOpt}--\cite{contreraspena2}.
 Stochastic volatility models and the multi-asset Black--Scholes model can also be thought of as constrained quantum systems \cite{contrerasSochasVolty, contrerasBusta}.
There are also gauge theoretical descriptions of the arbitrage process \cite{ilinski}.\\ \\
In \cite{contrerasFIRST}, inspired by the ideas in \cite{ilinski}, a generalization of the Black--Scholes (B--S) model that incorporates market imperfections through the presence of arbitrage bubbles was proposed.
In this case, the Black--Scholes equation is given by  
\begin{equation}
\frac{\partial \pi}{\partial t}+\frac{1}{2} \sigma^{2} S^{2} \frac{\partial^{2} \pi}{\partial S^{2}}+r \frac{\left(\sigma-\frac{\alpha f(S,t)}{r}\right)}{(\sigma-f(S,t))}\left(S \frac{\partial \pi}{\partial S}-\pi\right) = 0,
\end{equation}
where the function $f=f(S, t)$ is called the amplitude of the arbitrage bubble.
The above equation can written  as
\begin{equation} \label{BSEinteraction}
\frac{\partial \pi}{\partial t}+\frac{1}{2} \sigma^{2} S^{2} \frac{\partial^{2} \pi}{\partial S^{2}}+r\left(S \frac{\partial \pi}{\partial S}-\pi\right) + v(S,t) \left(S \frac{\partial \pi}{\partial S}-\pi\right)=0,
\end{equation}
with
\begin{equation} \label{potentialSt}
v(S,t) = \frac{(r-\alpha) f(S, t)}{\sigma-f(S, t)},
\end{equation}
which can be interpreted as the potential of an external time-dependent force generated by the arbitrage bubble $f(S,t)$.
In fact, from a physicist's point of view, $v(S,t)$ is equivalent to a magnetic and electric field of the same strengths.
Thus, market imperfections produce interactions that are equivalent to an external electromagnetic field of force that acts on the Black--Scholes particle.\\
By using the coordinate transformation 
\begin{equation} \label{transf1}
x=\ln S-\left(r-\frac{1}{2} \sigma^{2}\right) t,
\end{equation}
the interacting Black--Scholes equation (\ref{BSEinteraction}) becomes
\begin{equation} \label{BSExinteraction2}
\frac{\partial \pi}{\partial t}+\frac{1}{2} \sigma^{2} \frac{\partial^{2} \pi}{\partial x^{2}}-r \pi+\frac{(r-\alpha) \tilde{f}}{\sigma-\check{f}}\left(\frac{\partial \pi}{\partial x}-\pi\right) = 0,
\end{equation}
where
\begin{equation}
\check{f}(x, t)=f\left(\mathrm{e}^{x+\left(r-\frac{1}{2} \sigma^{2}\right) t}, t\right).
\end{equation}
Now by defining
\begin{equation} \label{transf3}
\pi(x, t)=\mathrm{e}^{-r(T-t)} \psi(x, t),
\end{equation}
equation (\ref{BSExinteraction2}) becomes
\begin{equation} \label{BSExinteraction3}
\frac{\partial \psi(x, t)}{\partial t}+\frac{1}{2} \sigma^{2} \frac{\partial^{2} \psi(x, t)}{\partial x^{2}}+v(x, t)\left(\frac{\partial \psi(x, t)}{\partial x}-\psi(x, t)\right) = 0,
\end{equation}
with
\begin{equation} 
v(x, t)=\frac{(r-\alpha) \check{f}(x, t)}{\sigma-\check{f}(x, t)}.
\end{equation}
Now, finally by doing a Wick rotation in time
\begin{equation} \label{WICKROTATION}
t=-i \tau,
\end{equation}
equation (\ref{BSExinteraction3}) becomes
\begin{equation} \label{SCHROinteraction}
i \frac{\partial \psi(x, \tau)}{\partial \tau} = - \frac{1}{2} \sigma^{2} \frac{\partial^{2} \psi(x, \tau)}{\partial x^{2}} - v(x, \tau)\left(\frac{\partial \psi(x, \tau)}{\partial x}-\psi(x, \tau)\right).
\end{equation}
The last equation is an interacting Schr\"{o}dinger equation for a particle of mass $1 / \sigma^{2}$ with wave function $\psi(x, t)$ in an external time-dependent field generated by the potential $v(x, t)$.
If we write the above interacting Schr\"{o}dinger equation as
\begin{equation}
\frac{\partial \psi(x, t)}{\partial t}=\check{H} \psi(x, t)
\end{equation}
one can express the Hamiltonian operator as
\begin{equation} \label{Hxinteraction3}
\check{H}=-\frac{1}{2} \sigma^{2} \frac{\partial^{2}}{\partial x^{2}}-v(x, t)\left(\frac{\partial}{\partial x}-I\right).
\end{equation}
Note that if $f = 0$, the potential $v =0$ and so (\ref{Hxinteraction3})
corresponds to the Hamiltonian of a free particle.
When the amplitude $f$ of the bubble is small, and so the second term in (\ref{Hxinteraction3}) can be thought of as a perturbation of the free Hamiltonian.
Since $f$ is time-dependent, one can apply the interaction picture to study the effect of the perturbation on the free solution, that is, on the usual Black--Scholes solution.\\
In the following section I develop these ideas but not in the $(x, \tau)$ space, but directly in the $(S, t)$ space.

\section{The Euclidean interaction picture and option pricing}
Consider again the Black--Scholes equation (\ref{BSEinteraction}) in the $(S,t)$ space in the presence of an arbitrage bubble $f(S,t)$.
This equation must be integrated with the final condition
\begin{equation}
\pi(S,T) = \Phi(S).
\end{equation}
The function $\Phi$ is called the contract function and defines the type of option.
Note that equation (\ref{BSEinteraction}) must be integrated backward in time from the future time $t=T$ to the present time $t=0$.
One can change the direction of time by using the change of variables given by
\begin{equation}
\tau = T - t
\end{equation}
which implies that
\begin{equation}
\frac{\partial }{\partial \tau}  = - \frac{\partial}{\partial t}
\end{equation}
so (\ref{BSEinteraction}) can be written as forward $\tau$ time Euclidean Schr\"{o}dinger like equation 
\begin{equation} \label{BSEinteractionv3}
\frac{\partial \pi}{\partial \tau} = \frac{1}{2} \sigma^{2} S^{2} \frac{\partial^{2} \pi}{\partial S^{2}}+r\left(S \frac{\partial \pi}{\partial S}-\pi\right) + v(S, \tau) \left(S \frac{\partial \pi}{\partial S}-\pi\right).
\end{equation}
From (\ref{BSEinteractionv3}) one can identify the Euclidean Hamiltonian operator as
\begin{equation} 
\check{H} = \frac{1}{2} \sigma^{2} \check{T} + r \check{P} + v(S, \tau) \check{P},
\end{equation}
with
\begin{equation} 
\check{T} = S^{2} \frac{\partial^{2} }{\partial S^{2}},
\end{equation}
and
\begin{equation} 
\check{P} = \left(S \frac{\partial \pi}{\partial S} - \check{I} \right).
\end{equation}
When the amplitude of the bubble is zero, the potential function $v(S, \tau)$ also is zero, and the Hamiltonian reduces to
\begin{equation} 
	\check{H}_0 = \frac{1}{2} \sigma^{2} \check{T} + r \check{P}, 
\end{equation}
which gives the evolution of the usual Black--Scholes model.
Thus one has that
\begin{equation} 
	\check{H} = \check{H}_0 + \check{V},
\end{equation}
with
\begin{equation} 
\check{V} = \check{V}(S, \tau) = v(S, \tau) \check{P}.
\end{equation}
In this way,  the interacting Black--Scholes equation is
\begin{equation} \label{BSEinteractionv4}
\frac{\partial \pi}{\partial \tau} = \check{H}_0 \pi + \check{V} \pi 
\end{equation}
Now introduce the Euclidean interaction picture by defining the interaction option price
\begin{equation} \label{interaction_pi_0}
\pi_I(S, \tau) = e^{ -\check{H}_0 \tau} \pi(S, \tau),
\end{equation}
or
\begin{equation} \label{interaction_pi}
\pi(S, \tau) = e^{ \check{H}_0 \tau} \pi_I (S, \tau).
\end{equation}
Taking the derivative
with respect to time,
\begin{equation}
\frac{\partial \pi(S, \tau)}{\partial \tau} =  \check{H}_0 e^{ \check{H}_0 \tau} \pi_I (S, \tau) + e^{ \check{H}_0 \tau}  \frac{\partial \pi_I(S, \tau)}{\partial \tau}.
\end{equation}
By substituting the above equation in (\ref{BSEinteractionv4})  and using (\ref{interaction_pi}), one obtains
\begin{equation}
\check{H}_0 e^{ \check{H}_0 \tau} \pi_I (S, \tau) + e^{ \check{H}_0 \tau}  \frac{\partial \pi_I(S, \tau)}{\partial \tau} = \check{H}_0 e^{ \check{H}_0 \tau} \pi_I (S, \tau) + \check{V} e^{ \check{H}_0 \tau} \pi_I (S, \tau),
\end{equation}
or
\begin{equation} \label{optionpricediracevolution}
\frac{\partial \pi_I(S, \tau)}{\partial \tau} = e^{ - \check{H}_0 \tau} \check{V}(S, \tau) e^{ \check{H}_0 \tau} \pi_I (S, \tau).
\end{equation}
If $\check{O}(S,\tau)$ is an operator defined in the Sch\"{o}dinger picture, then in the interaction picture it is given by
\begin{equation}
\check{O}_I (S,\tau) =  e^{ - \check{H}_0 \tau} \check{O}(S, \tau) e^{ \check{H}_0 \tau},
\end{equation}
and one can verify that
\begin{equation}
\frac{\partial \check{O}_I (S,\tau)}{\partial \tau}  =  [ \check{H}_0, \check{O}_I (S,\tau)] + e^{ - \check{H}_0 \tau} \frac{\partial \check{O} (S,\tau)}{\partial \tau} e^{ \check{H}_0 \tau}.
\end{equation}
Finally, equation (\ref{optionpricediracevolution}) can be written as
\begin{equation} \label{optionpricediracevolutionv2}
\frac{\partial \pi_I(S, \tau)}{\partial \tau} = \check{V}_I (S,\tau) \pi_I (S, \tau),
\end{equation}
with
\begin{equation} \label{potentialoperatorIP}
\check{V}_I (S,\tau)  =  e^{ - \check{H}_0 \tau}  \check{V} (S,\tau) e^{ \check{H}_0 \tau} =  e^{ - \check{H}_0 \tau}  v(S, \tau) \check{P} e^{ \check{H}_0 \tau}.
\end{equation}
Equation (\ref{optionpricediracevolutionv2}) can be integrated to give
\begin{equation} \label{optionpricediracevolutionv3}
\pi_I(S, \tau) = \pi_I(S, \tau_0) + \int_{\tau_0}^{\tau} \check{V}_I (S, \tau') \pi_I (S, \tau ') \ d \tau'.
\end{equation}
By choosing $\tau_0 = 0$ and using the fact that
\begin{equation} 
\pi_I(S, 0) = e^{ -\check{H}_0 \ 0} \pi(S, 0) = \Phi(S) 
\end{equation}
one has finally that
\begin{equation} \label{optionpricediracevolutionv3b}
\pi_I(S, \tau) = \Phi(S)  + \int_{0}^{\tau} \check{V}_I (S, \tau') \pi_I (S, \tau ') \ d \tau'.
\end{equation}
One can apply (\ref{optionpricediracevolutionv3b}) to construct an approximate solution for the option price in presence of a perturbation generated by an arbitrage bubble.

\section{The square bubble}
In that follows, I consider an important special case: the square-time bubble defined by  
\begin{equation}
f(t)=\left\{\begin{array}{ll}
0 & t<T_{1} \\
f_0 & T_{1}<t<T_{2} \\
0 & T_{2}<t<T
\end{array}\right.
\end{equation}
or, in terms of $\tau$,
\begin{equation}
f(\tau)=\left\{\begin{array}{ll}
0 & \tau < \tau_{1} \\
f_0 & \tau_{1}< \tau < \tau_{2} \\
0 & \tau_{2}< \tau < T
\end{array}\right.
\end{equation}
where $\tau_1 = T - T_2$ and $\tau_2 = T - T_1$.
This bubble is important, because any other shape of a bubble can be approximated locally by the square case.
Because this bubble is independent of the $S$ coordinate, the potential function (\ref{potentialSt}) is also independent of $S$, and given by
\begin{equation} \label{potentialfunction2}
v(\tau)=\left\{\begin{array}{ll}
0 & \tau < \tau_{1} \\
v_0 = \frac{(r-\alpha) f_0}{\sigma-f_0}, & \tau_{1}< \tau < \tau_{2} \\
0 & \tau_{2}< \tau < T
\end{array}\right.
\end{equation}
and the potential operator (\ref{potentialoperatorIP}) becomes
\begin{equation} \label{potentialoperatorIP2}
\check{V}_I (S,\tau)  = v(\tau) e^{ - \check{H}_0 \tau}  \check{P} e^{ \check{H}_0 \tau}.
\end{equation}
To evaluate (\ref{potentialoperatorIP2}) one need the commutation relations between $\check{H}_0$ and $\check{P}$.
In fact
\begin{equation}
[\check{H}_0, \check{P} ] = [\frac{1}{2} \sigma^{2} \check{T} + r \check{P}, \check{P} ] = \frac{1}{2} \sigma^{2}  [ \check{T}, \check{P} ].
\end{equation}
Now
\begin{eqnarray}
[ \check{T}, \check{P} ]  = & \check{T} \check{P}  - \check{P} \check{T} \\
  = & S^{2} \frac{\partial^{2} }{\partial S^{2}}  \left(S \frac{\partial }{\partial S} - \check{I} \right)  - \left(S \frac{\partial }{\partial S} - \check{I} \right) \left( S^{2} \frac{\partial^{2} }{\partial S^{2}}  \right) \\
 = & S^{2} \frac{\partial}{\partial S}  \left(  \frac{\partial }{\partial S} + S \frac{\partial^{2} }{\partial S^{2}}  - \frac{\partial }{\partial S}   \right) - \left( 
 S  \frac{\partial}{\partial S} \left( S^{2} \frac{\partial^{2}  }{\partial S^{2}}  \right) -  S^{2} \frac{\partial^{2}  }{\partial S^{2}}   \right)    \\ 
= &  S^{2} \left( \frac{\partial^{2} }{\partial S^{2}}  +  S \frac{\partial^{3} }{\partial S^{3}}   \right) - \left( 
S  \left( 2 S \frac{\partial^{2} }{\partial S^{2}} + S^2 \frac{\partial^{3}  }{\partial S^{3}}  \right) -  S^{2} \frac{\partial^{2}  }{\partial S^{2}}   \right)   \\  
= &   S^{2} \left( \frac{\partial^{2}}{\partial S^{2}}  +  S \frac{\partial^{3} }{\partial S^{3}}   \right) - \left( 
S^2  \frac{\partial^{2}  }{\partial S^{2}} + S^3 \frac{\partial^{3}  }{\partial S^{3}}  \right)  \\ 
= & 0,
\end{eqnarray}
thus $\check{P}$ commutes with the free Hamiltonian $\check{H}_0$, so (\ref{potentialoperatorIP2}) is equal to
\begin{equation} \label{potentialoperatorIP3}
\check{V}_I (S,\tau)  = v(\tau) \check{P} e^{ - \check{H}_0 \tau}   e^{ \check{H}_0 \tau} = v(\tau) \check{P},
\end{equation}
and  (\ref{optionpricediracevolutionv3b}) becomes
\begin{equation} \label{optionpricediracevolutionv4}
\pi_I(S, \tau) = \Phi(S) + \int_{0}^{\tau} v(\tau') \check{P} \ \pi_I (S, \tau ') \ d \tau'.
\end{equation}
Note that for $0 < \tau < \tau_{1} $, using (\ref{potentialfunction2}) shows that $v(\tau) = 0$, so in this temporal interval the option price is constant and equal to its initial value
\begin{equation}
\pi_I(S, \tau) = \Phi(S).
\end{equation}
For $ \tau_{1} < \tau < \tau_{2} $ the option price is given by
\begin{equation} \label{optionpricediracevolutionv5}
\pi_I(S, \tau) = \Phi(S) + \int_{\tau_1}^{\tau} v(\tau') \check{P} \ \pi_I (S, \tau ') \ d \tau'.
\end{equation}
At last, for $ \tau_{2} < \tau < T $, the potential function $v(\tau') = 0 $ again, so the option price is again constant and so must be equal to its value at $\tau = \tau_2$
\begin{equation} 
\pi_I(S, \tau_2) = \Phi(S) + \int_{\tau_1}^{\tau_2} v(\tau') \check{P} \ \pi_I (S, \tau ') \ d \tau', \ \ \ \ \  \  \ \   \tau_{1} < \tau < \tau_{2}.
\end{equation}
Thus, by equation (\ref{potentialfunction2}) one has that
\begin{equation} \label{solutionsquarebubble}
\pi_I(S, \tau) =\left\{\begin{array}{ll}
\Phi(S) & 0 < \tau < \tau_{1} \\
\Phi(S) + v_0 \int_{\tau_1}^{\tau} \check{P} \ \pi_I (S, \tau ') \ d \tau' & \tau_{1}< \tau < \tau_{2} \\
\Phi(S) + v_0 \int_{\tau_1}^{\tau_2} \check{P} \ \pi_I (S, \tau ') \ d \tau' & \tau_{2}< \tau < T
\end{array}\right.
\end{equation}
By taken the derivative with respect to time of the above equation one has that 
\begin{equation}
\frac{\partial \pi_I(S, \tau) }{\partial \tau } =\left\{\begin{array}{ll}
0 & 0 < \tau < \tau_{1} \\
 v_0  \check{P} \ \pi_I (S, \tau) & \tau_{1}< \tau < \tau_{2} \\
0  & \tau_{2}< \tau < T
\end{array}\right.
\end{equation}
So for $\tau_{1}< \tau < \tau_{2}$
\begin{equation}
\frac{\partial \pi_I(S, \tau) }{\partial \tau } = v_0  \check{P} \ \pi_I (S, \tau),
\end{equation}
and since the right side does not depend on time and using the initial condition $\pi_I(S, \tau_1) = \Phi(S)$ it can be  integrated to give 
\begin{equation}
\pi_I(S, \tau)  = e^{v_0  \check{P} (\tau - \tau_1)} \Phi(S)
\end{equation}
so 
\begin{equation}
\pi_I(S, \tau_2)  = e^{v_0  \check{P} (\tau_2 - \tau_1)} \Phi(S)
\end{equation}
and (\ref{solutionsquarebubble}) can be written also as
\begin{equation} \label{solutionsquarebubble2}
\pi_I(S, \tau) =\left\{\begin{array}{ll}
\Phi(S) & 0 < \tau < \tau_{1} \\
e^{v_0  \check{P} (\tau - \tau_1)} \Phi(S)& \tau_{1}< \tau < \tau_{2} \\
 e^{v_0  \check{P} (\tau_2 - \tau_1)} \Phi(S) & \tau_{2}< \tau < T.
\end{array}\right.
\end{equation}
The evolution of the option price in the Schr\"{o}dinger picture is then given by (\ref{interaction_pi}) 
\begin{equation} \label{solutionsquarebubble3}
\pi(S, \tau) =\left\{\begin{array}{ll}
e^{ \check{H}_0 \tau}  \Phi(S) & 0 < \tau < \tau_{1} \\
e^{ \check{H}_0 \tau}  e^{v_0  \check{P} (\tau - \tau_1)} \Phi(S)& \tau_{1}< \tau < \tau_{2} \\
e^{ \check{H}_0 \tau}  e^{v_0  \check{P} (\tau_2 - \tau_1)} \Phi(S) & \tau_{2}< \tau < T
\end{array}\right.
\end{equation}
Due to the fact that $\check{H}_0$ commutes with $\check{P}$ for the square bubble, 
\begin{equation} \label{solutionsquarebubble4}
\pi(S, \tau) =\left\{\begin{array}{ll}
e^{ \check{H}_0 \tau}  \Phi(S) & 0 < \tau < \tau_{1} \\
e^{ \check{H}_0 \tau + v_0  \check{P} (\tau - \tau_1)} \Phi(S)& \tau_{1}< \tau < \tau_{2} \\
e^{  \check{H}_0 \tau + v_0  \check{P} (\tau_2 - \tau_1)} \Phi(S) & \tau_{2}< \tau < T
\end{array}\right.
\end{equation}
which is the same as
\begin{equation} \label{solutionsquarebubble5}
\pi(S, \tau) =\left\{\begin{array}{ll}
e^{ \check{H}_0 \tau}  \Phi(S) & 0 < \tau < \tau_{1} \\   
e^{ \left[\check{H}_0 + v_0  \check{P}  \right] (\tau - \tau_1)  }  e^{ \check{H}_0 \tau_1}  \Phi(S)& \tau_{1}< \tau < \tau_{2} \\
e^{ \check{H}_0 (\tau - \tau_2) } e^{ \left[ \check{H}_0 + v_0  \check{P} \right] (\tau_2 - \tau_1) }  e^{ \check{H}_0 \tau_1}  \Phi(S) & \tau_{2}< \tau < T
\end{array}\right.
\end{equation}
Now
\begin{equation} 
\check{H}_0 + v_0  \check{P}  = \frac{1}{2} \sigma^{2} \check{T} + r \check{P} +  v_0  \check{P} = \frac{1}{2} \sigma^{2} \check{T} + (r + v_0) \check{P} = \frac{1}{2} \sigma^{2} \check{T} + \bar{r} \check{P},
\end{equation}
where $\bar{r} = r + v_0 $ is a `dressed' interest rate.
One can see that the total Hamiltonian is the same as the free Balck--Scholes Hamiltonian but with the `bare' interest rate $r$ replaced by the effective one $\bar{r}$.
If one defines the free Hamiltonian  as $\check{H}_0^{\bar{r}}  = \frac{1}{2} \sigma^{2} \check{T} + \bar{r} \check{P}$,
then
\begin{equation} 
\pi(S, \tau) =\left\{\begin{array}{ll}
e^{ \check{H}_0 \tau}  \Phi(S) & 0 < \tau < \tau_{1} \\   
e^{ \check{H}_0^{\bar{r}}  (\tau - \tau_1)  }  e^{ \check{H}_0 \tau_1}  \Phi(S)& \tau_{1}< \tau < \tau_{2} \\
e^{ \check{H}_0 (\tau - \tau_2) } e^{ \check{H}_0^{\bar{r}} (\tau_2 - \tau_1) }  e^{ \check{H}_0 \tau_1}  \Phi(S) & \tau_{2}< \tau < T
\end{array}\right.
\end{equation}
In this way, by using the interaction picture one can obtain a formal solution of the problem.
For example, for a European call $ \Phi(S) = max\{ S-K, 0\} $, the first part $ 0 < \tau < \tau_{1}$ of the solution given by
\begin{equation} 
\pi(S, \tau) = e^{ \check{H}_0 \tau} \Phi(S)
\end{equation} 
corresponds to the usual closed form Black--Scholes solution.
However, for $\tau_{1}< \tau < \tau_{2}$ or $\tau_{2}< \tau < T$, though the evolution is given by a free Black--Scholes Hamiltonian, the solution is not given by a usual solution to the Black--Scholes equation.
This is because the initial conditions are not given by $ \Phi(S) = max\{ S-K, 0\} $ but instead by $\Phi_1(S) = e^{ \check{H}_0 \tau_1}  \Phi(S) $ and $\Phi_2(S) = e^{ \check{H}_0^{\bar{r}} (\tau_2 - \tau_1) }  e^{ \check{H}_0 \tau_1}  \Phi(S) $ respectively. For this reason, obtaining an explicit formula for the solution in these time intervals is a non trivial task.
It can only be accomplished  numerically.

\section{The first terms of the perturbation solution}
Consider equation (\ref{solutionsquarebubble}) for the intermediate region of time $\tau_1 < \tau < \tau_2$:
\begin{equation} 
\pi_I (S, \tau) = \Phi(S) + v_0 \int_{\tau_1}^{\tau} \check{P} \ \pi_I (S, \tau ') \ d \tau'. 
\end{equation}
By iterating this equation one obtains 
\begin{equation} 
\begin{array} {ll}
\pi_I (S, \tau) = & \Phi(S) + v_0 \int_{\tau_1}^{\tau} \check{P} \  \Phi(S) \ d \tau' +  \\
& v_0 \int_{\tau_1}^{\tau} \check{P} \  \left(   v_0 \int_{\tau_1}^{\tau'} \check{P} \  \Phi(S) \ d \tau'' \right) \ d \tau' +  \\ 
&  v_0 \int_{\tau_1}^{\tau} \check{P} \  \left(   v_0 \int_{\tau_1}^{\tau'} \check{P} \  \left( v_0 \int_{\tau_1}^{\tau''} \check{P} \  \Phi(S) \ d \tau''' \right) \ d \tau'' \right) \ d \tau' +  \cdots,   
\end{array}
\end{equation}
that is,
\begin{equation} 
\begin{array}{ll}
\pi_I (S, \tau) = & \Phi(S) + v_0 \ \check{P} \  \Phi(S) \ \left( \int_{\tau_1}^{\tau} d \tau' \right) +  \\
& v_0^2 \ \check{P} \ \check{P} \  \Phi(S) \  \left( \int_{\tau_1}^{\tau}  \int_{\tau_1}^{\tau'} d \tau''  \ d \tau' \right) +  \\ 
&  v_0^3 \ \check{P} \ \check{P} \ \check{P} \ \Phi(S) 
\left( \int_{\tau_1}^{\tau} \int_{\tau_1}^{\tau'} \int_{\tau_1}^{\tau''} \ d \tau'''   \ d \tau''  \ d \tau'  \right) +  \cdots.    
\end{array}
\end{equation}
Integrating with respect to time gives
\begin{equation}
\int_{\tau_1}^{\tau} d \tau' = \tau - \tau_{1}
\end{equation}
\begin{equation}
\int_{\tau_1}^{\tau}  \int_{\tau_1}^{\tau'} d \tau''  \ d \tau'  = \int_{\tau_1}^{\tau} \left( \tau' - \tau_{1} \right) \ d \tau' = \frac{1}{2} \left( \tau - \tau_{1}   \right)^2
\end{equation}
\begin{equation}
\int_{\tau_1}^{\tau} \int_{\tau_1}^{\tau'} \int_{\tau_1}^{\tau''} \ d \tau'''   \ d \tau''  \ d \tau'   =  \int_{\tau_1}^{\tau} \frac{1}{2} \left( \tau' - \tau_{1}   \right)^2 \ d \tau' = \frac{1}{3!} \left( \tau - \tau_{1}   \right)^3
\end{equation}
\begin{equation*}
\vdots
\end{equation*}
and so
\begin{eqnarray*}
\pi_I (S, \tau) = & \Phi(S) + v_0 \ \check{P} \  \Phi(S) \ \frac{1}{1!} \left( \tau - \tau_{1}   \right) +  \\
& v_0^2 \ \check{P} \ \check{P} \  \Phi(S) \  \frac{1}{2!} \left( \tau - \tau_{1}   \right)^2 +  \\ 
&  v_0^3 \ \check{P} \ \check{P} \ \check{P} \ \Phi(S) 
\frac{1}{3!} \left( \tau - \tau_{1}   \right)^3 +  \cdots   
\end{eqnarray*}
or
\begin{eqnarray*} \label{threefirstterms}
\pi_I (S, \tau) = & \Big( \check{I} + \frac{1}{1!} v_0 \ \check{P}  \left( \tau - \tau_{1}   \right) +  \\
&  \frac{1}{2!} v_0^2 \ \check{P}^2 \left( \tau - \tau_{1}   \right)^2 +  \\ 
& \frac{1}{3!}  v_0^3 \ \check{P}^3 \left( \tau - \tau_{1}   \right)^3 +  \cdots  \Big) \Phi(S)  
\end{eqnarray*} 
Now by kipping only the first four terms in (\ref{threefirstterms}) one has
\begin{equation}
\pi_I (S, \tau) \approx \Big( \check{I} + \frac{1}{1!}  v_0 \left( \tau - \tau_{1}   \right)  \ \check{P}   +  \frac{1}{2!} v_0^2 \left( \tau - \tau_{1}   \right)^2  \ \check{P}^2 +
\frac{1}{3!} v_0^3 \left( \tau - \tau_{1}   \right)^3  \ \check{P}^3 \Big) \Phi(S).
\end{equation}
Then the option price in the Schr\"{o}dinger picture is
\begin{equation}
\pi (S, \tau) \approx e^{ \check{H}_0 \tau}  \Big( \check{I} + \frac{1}{1!}  v_0 \left( \tau - \tau_{1}   \right)  \ \check{P}   +  \\
\frac{1}{2!} v_0^2 \left( \tau - \tau_{1}   \right)^2  \ \check{P}^2 
+
\frac{1}{3!} v_0^3 \left( \tau - \tau_{1}   \right)^3  \ \check{P}^3
\Big) \Phi(S),
\end{equation}
but as $\check{H}_0$ commutes with $\check{P}$ one has that
\begin{equation}
\pi (S, \tau) \approx  \Big( \check{I} + \frac{1}{1!}  v_0 \left( \tau - \tau_{1}   \right)  \ \check{P}   +  \frac{1}{2!} v_0^2 \left( \tau - \tau_{1}   \right)^2  \ \check{P}^2 
+
\frac{1}{3!} v_0^3 \left( \tau - \tau_{1}   \right)^3  \ \check{P}^3
\Big)  e^{ \check{H}_0 \tau}  \Phi(S).
\end{equation}
But $e^{ \check{H}_0 \tau}  \Phi(S) $  is just the Call $C(S,t)$ solution at time $\tau$ so
\begin{equation}
C(S, \tau) = e^{ \check{H}_0 \tau}  \Phi(S)
\end{equation}
and then
\begin{equation} \label{optionprice2aprox}
\pi (S, \tau) \approx C(S, \tau) + v_0 \left( \tau - \tau_{1}   \right)  \ \check{P} \ C(S,\tau)  +  \frac{1}{2} v_0^2 \left( \tau - \tau_{1}   \right)^2  \ \check{P}^2 \ C(S, \tau) +
\frac{1}{3!} v_0^3 \left( \tau - \tau_{1}   \right)^3  \ \check{P}^3 \ C(S, \tau).
\end{equation}
Now 
\begin{equation}
\check{P} \ C(S, \tau) = \left(  S \frac{\partial  }{\partial S} - \check{I} \right) C(S, \tau) =  S \frac{\partial C(S, \tau)  }{\partial S} - C(S, \tau).
\end{equation}
The derivative of a Call is called Delta
\begin{equation}
\Delta(S, \tau) = \frac{\partial C(S, \tau)  }{\partial S}
\end{equation}
so
\begin{equation} \label{PCall}
\check{P} \ C(S, \tau) = S \ \Delta(S, \tau) - C(S, \tau).
\end{equation}
For  $\check{P}^2$ one has
\begin{eqnarray}
\check{P}^2 = & \left(  S \frac{\partial  }{\partial S} - \check{I} \right) \left(  S \frac{\partial  }{\partial S} - \check{I} \right) \\
= & S \frac{\partial  }{\partial S} \left(  S \frac{\partial  }{\partial S} - \check{I}    \right) - \left(  S \frac{\partial  }{\partial S} - \check{I} \right)     \\
= &   S \frac{\partial  }{\partial S} \left(  S \frac{\partial  }{\partial S}  \right) -   S \frac{\partial  }{\partial S}  - \left(  S \frac{\partial  }{\partial S} - \check{I} \right)  \\
= & S \left( \frac{\partial  }{\partial S} + S \frac{\partial^2  }{\partial S^2} \right) - 2 S \frac{\partial  }{\partial S} + \check{I}    \\
= &   S^2 \frac{\partial^2  }{\partial S^2} - S \frac{\partial  }{\partial S} +  \check{I}
\end{eqnarray}
and so 
\begin{equation}
\check{P}^2 \ C(S, \tau) = S^2 \ \frac{\partial^2 C(S, \tau) }{\partial S^2} - S \ \frac{\partial C(S, \tau)  }{\partial S} +  C(S, \tau).
\end{equation}
The second derivative of a Call is called Gamma:
\begin{equation}
\Gamma(S, \tau) = \frac{\partial^2 C(S, \tau) }{\partial S^2},
\end{equation}
so
\begin{equation} \label{P2Call}
\check{P}^2 C(S, \tau) = S^2 \ \Gamma(S, \tau) - S \ \Delta(S, \tau) +  C(S, \tau).
\end{equation}
Lastly, for $\check{P}^3$ one has
\begin{eqnarray}
\check{P}^3 = & \left(  S \frac{\partial  }{\partial S} - \check{I} \right) \left(  S \frac{\partial  }{\partial S} - \check{I} \right)^2 \\
= & \left(  S \frac{\partial  }{\partial S} - \check{I} \right)
\left(  S^2 \frac{\partial^2  }{\partial S^2} - S \frac{\partial  }{\partial S} +  \check{I} \right) \\
= &   S \frac{\partial  }{\partial S} \left( S^2 \frac{\partial^2  }{\partial S^2} - S \frac{\partial  }{\partial S}  \right) + S \frac{\partial  }{\partial S} - \left( S^2 \frac{\partial^2  }{\partial S^2} - S \frac{\partial  }{\partial S} +  \check{I} \right) \\
= &  S \left( 2 S \frac{\partial^2  }{\partial S^2 }  + S^2  \frac{\partial^3  }{\partial S^3 } - \frac{\partial  }{\partial S} - S \frac{\partial^2  }{\partial S^2}  \right) + S \frac{\partial  }{\partial S} -  S^2 \frac{\partial^2  }{\partial S^2} + S \frac{\partial  }{\partial S} -  \check{I}  \\
= &    S^3  \frac{\partial^3  }{\partial S^3 } + S \frac{\partial  }{\partial S} - \check{I},
\end{eqnarray}
and so
\begin{equation}
\check{P}^3 \ C(S, \tau) = S^3 \ \frac{\partial^3 C(S, \tau) }{\partial S^3 } + S \ \frac{\partial C(S, \tau) }{\partial S} - C(S, \tau).
\end{equation}
The third derivative of a Call is called the Speed:
\begin{equation}
Spd(S, \tau) = \frac{\partial^3 C(S, \tau) }{\partial S^3 },
\end{equation}
thus
\begin{equation} \label{P3Call}
\check{P}^3 \ C(S, \tau) = S^3 \ Spd(S, \tau) + S \ \Delta(S, \tau) - C(S, \tau).
\end{equation} \\
By substituting (\ref{PCall}), (\ref{P2Call}) and (\ref{P3Call}) in (\ref{optionprice2aprox}) we obtain \\
\begin{equation} 
\begin{array} {lll}
\pi (S, \tau) \approx &  C(S, \tau) + v_0 \left( \tau - \tau_{1} \right) \left( S \ \Delta(S, \tau) - C(S, \tau)  \right)  + \\
&   \frac{1}{2} v_0^2 \left( \tau - \tau_{1}   \right)^2 \left( S^2 \ \Gamma(S, \tau) - S \ \Delta(S, \tau) +  C(S, \tau)    \right)   +      \\
& \frac{1}{3!} v_0^3 \left( \tau - \tau_{1}   \right)^3 \left( S^3 \ Spd(S, \tau) + S \ \Delta(S, \tau) - C(S, \tau)
\right),        
\end{array}
\end{equation} 
or
\begin{equation} \label{naiveapproxsolution}
\begin{array} {lll}
\pi (S, \tau) \approx & \left( 1 - v_0 \left( \tau - \tau_{1} \right) + \frac{1}{2} v_0^2 \left( \tau - \tau_{1}   \right)^2 - \frac{1}{3!} v_0^3 \left( \tau - \tau_{1}   \right)^3  \right)  C(S, \tau) + \\
& \left( v_0 \left( \tau - \tau_{1} \right) - \frac{1}{2} v_0^2 \left( \tau - \tau_{1}   \right)^2 + \frac{1}{3!} v_0^3 \left( \tau - \tau_{1}   \right)^3   \right)  S \ \Delta(S, \tau) + \\
& \frac{1}{2} v_0^2 \left( \tau - \tau_{1}   \right)^2  S^2 \ \Gamma(S, \tau) + \\
& \frac{1}{3!} v_0^3 \left( \tau - \tau_{1}   \right)^3   S^3 \ Spd(S, \tau).
\end{array} 
\end{equation} \\ \\
Figure 1 shows the behaviour of the approximate perturbation solution (\ref{naiveapproxsolution}) for several values of $f_0/\sigma$.
Note that near $\frac{f_0}{\sigma} \approx 1$, the approximate solution gives a good qualitative description of the resonance reported in \cite{contrerasresonance}.
\begin{figure}[h!]
	\centering
	\includegraphics[scale=2.1]{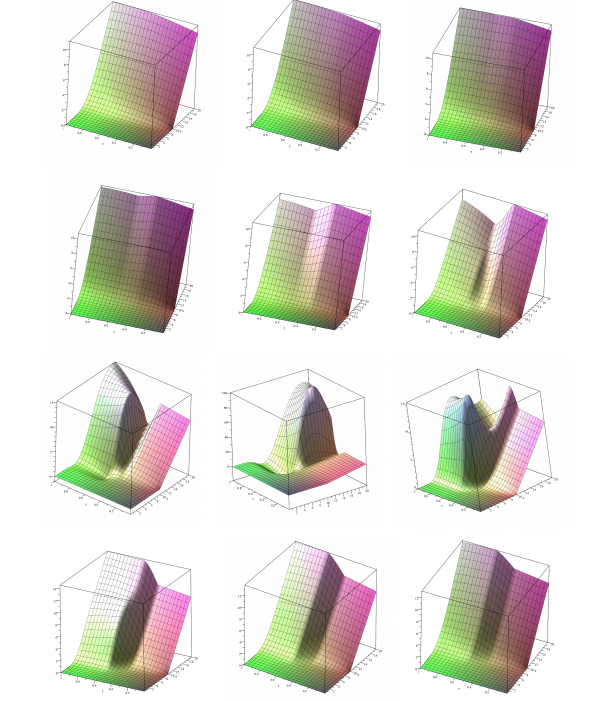}
	\caption{from left to right and from top to the bottom, approximate Call solution (\ref{naiveapproxsolution}) for $\frac{f_0}{\sigma}$ = 0, 0.1, 0.2,  0.4, 0.6, 0.8, 0.9, 0.95, 1.10, 1.20, 1.40, 1.80  respectively.}
\end{figure}
\noindent \\
\section{The exact $\pi(S, \tau)$ solution for the square bubble.}
In this section we compute the exact solution for the option price for the square bubble in terms of the Greeks.
For this, consider equation (\ref{solutionsquarebubble3}). Now, since $\check{H}_0$ commutes with $\check{P}$ one has that 
\begin{equation} 
\pi(S, \tau) =\left\{\begin{array}{ll}
e^{ \check{H}_0 \tau}  \Phi(S) & 0 < \tau < \tau_{1} \\
e^{v_0  \check{P} (\tau - \tau_1)} e^{ \check{H}_0 \tau} \Phi(S)& \tau_{1}< \tau < \tau_{2} \\
e^{v_0  \check{P} (\tau_2 - \tau_1)} e^{ \check{H}_0 \tau} \Phi(S) & \tau_{2}< \tau < T
\end{array}\right.
\end{equation}
But as $ C(S,t) = e^{ \check{H}_0 \tau} \Phi(S)$, it follows that
\begin{equation} 
\pi(S, \tau) =\left\{\begin{array}{ll}
C(S, \tau) & 0 < \tau < \tau_{1} \\
e^{v_0 (\tau - \tau_1)  \check{P} } \ C(S, \tau)& \tau_{1}< \tau < \tau_{2} \\
e^{v_0 (\tau_2 - \tau_1) \check{P} } \ C(S, \tau) & \tau_{2}< \tau < T
\end{array}\right.
\end{equation}
Now we write the operator  $ \check{P}$ as
\begin{equation}
\check{P} = \check{K} - \check{I}
\end{equation}
with $ \check{K} = S \frac{\partial }{\partial S}$.
Again, since $\check{K}$ commutes with the identity,
\begin{equation} 
e^{x \check{P} } = e^{x \left(  \check{K} - \check{I}  \right)  } =  e^{x \check{K}  } e^{- x \check{I} },
\end{equation}
and so
\begin{equation} 
e^{v_0 (\tau - \tau_1)  \check{P} } \ C(S, \tau) = e^{ - v_0 (\tau - \tau_1) }  e^{v_0 (\tau - \tau_1)  \check{K} } \ C(S, \tau).
\end{equation}
In this way the exact solution is
\begin{equation} \label{exactsolution}
\pi(S, \tau) =\left\{\begin{array}{ll}
C(S, \tau) & 0 < \tau < \tau_{1} \\
 e^{ - v_0 (\tau - \tau_1) }  e^{v_0 (\tau - \tau_1)  \check{K} } \ C(S, \tau)& \tau_{1}< \tau < \tau_{2} \\
 e^{ - v_0 (\tau_2 - \tau_1) }  e^{v_0 (\tau_2 - \tau_1)  \check{K} } \ C(S, \tau) & \tau_{2}< \tau < T
\end{array}\right.
\end{equation}
Now 
\begin{equation} 
e^{x \check{K} } = \check{I} + \frac{1}{1!} x \check{K} + \frac{1}{2!} x^2 \check{K}^2 + \cdots, 
\end{equation}
so to compute the solution one needs all the powers of the $\check{K}$ operator.

\section{The powers of $\check{K}$}
If $\check{D}$ denotes the differentiation operator $\frac{\partial }{\partial S}$,  then 
\begin{equation} 
\check{K} = S \check{D}.
\end{equation}
Now, one has that
\begin{eqnarray}
\check{K} \left(S^{n} \check{D}^{n}\right) = &  S \frac{\partial}{\partial S}\left(S^{n} \frac{\partial^{n}}{\partial S^{n}}\right) \\
= &S\left(n S^{n-1} \cdot \frac{\partial^{n}}{\partial S^{n}}+S^{n} \frac{\partial^{n+1}}{\partial S^{n+1}}\right) \\
= & n S^{n} \frac{\partial^{n}}{\partial S^{n}}+S^{n+1} \frac{\partial^{n+1}}{\partial S^{n+1}},  
\end{eqnarray}
that is,
\begin{equation} \label{KSnDn}
\check{K}  \left(S^{n} \check{D}^{n}\right) = n \ S^{n} \check{D}^{n} + S^{n+1} \check{D}^{n+1}.
\end{equation}
One can use (\ref{KSnDn}) to evaluate all the powers of $\check{K}$.
For example, 
\begin{equation}
\check{K}^2 =  \check{K} \check{K} = \check{K} \left(S \check{D} \right)  = S \check{D} + S^2 \check{D}^2   = \check{K} + S^2 \check{D}^2.
\end{equation}
Also,
\begin{eqnarray}
\check{K}^3  &  = \check{K} \check{K}^2 = \check{K} \left( \check{K} + S^2 \check{D}^2 \right) = \check{K}^2  +  \check{K} \left( S^2 \check{D}^2 \right)    \\
 & =   \left( \check{K} + S^2 \check{D}^2 \right)  + \left( 2 S^2 \check{D}^2 + S^3 \check{D}^3   \right).   
\end{eqnarray} 
\begin{equation}
\check{K}^3 = \check{K} + 3 S^2 \check{D}^2 +  S^3 \check{D}^3.
\end{equation}
In the same way, one has
\begin{equation}
\check{K}^4 = \check{K} + 7 S^2 \check{D}^2 +  6 S^3 \check{D}^3  + S^4 \check{D}^4,
\end{equation}
\begin{equation}
\check{K}^5 = \check{K} + 15 S^2 \check{D}^2 +  25 S^3 \check{D}^3  + 10 S^4 \check{D}^4 + S^5 \check{D}^5,
\end{equation}
\begin{equation}
\check{K}^6 = \check{K} + 31 S^2 \check{D}^2 +  90 S^3 \check{D}^3  + 65 S^4 \check{D}^4 + 15 S^5 \check{D}^5 + S^6 \check{D}^6
\end{equation}
and so on.
One can arrange the coefficients of expansion of $\check{K}^n$ in terms of $\check{K}$ and $\check{D}$ in a Pascal-like triangle in the following form.\\ \\ \\
\begin{tabular}{>{$n=}l<{$\hspace{12pt}}*{13}{c}}
	1 &&&&&&&1&&&&&&\\
	2 &&&&&&1&&1&&&&&\\
	3 &&&&&1&&3&&1&&&&\\
	4 &&&&1&&7&&6&&1&&&\\
	5 &&&1&&15&&25&&10&&1&&\\
	6 &&1&&31&&90&&65&&15&&1&\\
	7 &1&&63&&301&&350&&140&&21&&1 \\
\end{tabular} \\ \\ \\
Let $\alpha_{n, m} \ \ (n, m = 1, 2, 3, ...) $ be the coefficients in this triangle.
Here $n$ indicates the vertical position (from top to bottom)  and $m$ indicates the horizontal position (from left to the right).
For example  $ \ \alpha_{3, 2} = 3$,  $ \ \alpha_{5, 3} = 25$, $ \ \alpha_{6, 6} = 1$, etc.
The coefficients in the triangle can be determined recursively by the following relations:   $\alpha_{n, 1} = 1$,  $\alpha_{n, n} = 1$  and
\begin{equation}
\alpha_{n m} = m \alpha_{n-1, m} + \alpha_{n-1, m-1}.
\end{equation}
For example, $\alpha_{3, 2} = 2 \ \alpha_{2, 2} + \alpha_{2, 1}  = 2 \cdot 1+ 1 = 3$.
\\ \\
In this way, the powers of the operator $\check{K}$ can be calculated as
\begin{equation}
\check{K}^n = \check{K} + \sum_{j=2}^{n}  \alpha_{n, j} \ S^j \check{D}^j.
\end{equation}
For example,
\begin{equation}
\check{K}^2 = \check{K} +  \alpha_{2, 2} \ S^2 \check{D}^2
\end{equation}
\begin{equation}
\check{K}^3 = \check{K} +  \alpha_{3, 2} \ S^2 \check{D}^2 + \alpha_{3, 3} \ S^3 \check{D}^3
\end{equation}
\begin{equation}
\check{K}^4 = \check{K} + \alpha_{4, 2} \ S^2 \check{D}^2 + \alpha_{4, 3} \ S^3 \check{D}^3 + \alpha_{4, 4} \ S^4 \check{D}^4
\end{equation}
\begin{equation}
\check{K}^5 = \check{K} + \alpha_{5, 2} \ S^2 \check{D}^2 + \alpha_{5, 3} \ S^3 \check{D}^3 + \alpha_{5, 4} \ S^4 \check{D}^4 + \alpha_{5, 5} \ S^5 \check{D}^5.
\end{equation}
Thus, the operator 
\begin{equation}
e^{ x \check{K}} = \check{I} + \frac{1}{1 !} x \check{K} + \frac{1}{2!} x^2 \check{K}^2 +  \frac{1}{3!} x^3 \check{K}^3 +  \frac{1}{4!} x^4 \check{K}^4 +  \frac{1}{4!} x^4 \check{K}^4 + \cdots
\end{equation}
is then equal to
\begin{eqnarray}
e^{ x \check{K}}  = &   \check{I} +  \\
 &  \frac{1}{1 !} x \check{K} +  \\
 & \frac{1}{2!} x^2 \left(   \check{K} + \alpha_{2, 2} \ S^2 \check{D}^2 \right) +   \\
& \frac{1}{3!} x^3 \left( \check{K} +  \alpha_{3, 2} \ S^2 \check{D}^2 + \alpha_{3, 3} \ S^3 \check{D}^3     \right) +  \\
& \frac{1}{4!} x^4 \left( \check{K} + \alpha_{4, 2} \ S^2 \check{D}^2 + \alpha_{4, 3} \ S^3 \check{D}^3 + \alpha_{4, 4} \ S^4 \check{D}^4     \right) +  \\
& \frac{1}{5!} x^5 \left(  \check{K} + \alpha_{5, 2} \ S^2 \check{D}^2 + \alpha_{5, 3} \ S^3 \check{D}^3 + \alpha_{5, 4} \ S^4 \check{D}^4 + \alpha_{5, 5} \ S^5 \check{D}^5    \right)   + \cdots \\
\end{eqnarray} 
or
\begin{eqnarray}
e^{ x \check{K}} =  &   \check{I} + \\
& \left( \frac{1}{1!} x +  \frac{1}{2!} x^2 + \frac{1}{3!} x^3 + \frac{1}{4!} x^4 +  \frac{1}{5!} x^5 + \cdots  \right) \check{K} + \\
&  \left( \frac{1}{2!} x^2 \alpha_{2, 2} +  \frac{1}{3!} x^3 \alpha_{3, 2} + \frac{1}{4!} x^4 \alpha_{4, 2} +  \frac{1}{5!} x^5    \alpha_{5, 2} +    \cdots  \right)  S^2 \check{D}^2 +  \\
& \left( \frac{1}{3!} x^3 \alpha_{3, 3} + \frac{1}{4!} x^4 \alpha_{4, 3} +  \frac{1}{5!} x^5    \alpha_{5, 3} +    \cdots  \right)  S^3\check{D}^3+       \\
& \left( \frac{1}{4!} x^4 \alpha_{4, 4} + \frac{1}{5!} x^5    \alpha_{5, 4} +    \cdots  \right)  S^4 \check{D}^4 +  \\
& \left( \frac{1}{5!} x^5   \alpha_{5, 5} +  \cdots  \right)  S^5 \check{D}^5 + \cdots  .\\
\end{eqnarray} 
But
\begin{equation}
\frac{1}{1!} x +  \frac{1}{2!} x^2 + \frac{1}{3!} x^3 + \frac{1}{4!} x^4 +  \frac{1}{5!} x^5 + \cdots = e^x -1,
\end{equation}
and if one define the functions
\begin{equation}
Q_0 (x) = 1, 
\end{equation}
\begin{equation}
Q_1 (x) = e^x -1, 
\end{equation}
and
\begin{equation}
Q_j (x) = \sum_{m=j}^{\infty} \alpha_{m, j} \frac{x^m}{m!}   \ \ \  j = 2, 3, 4, \cdots
\end{equation}
then the exponential of $\check{K}$ is 
\begin{eqnarray*} \label{new_exponential}
e^{ x \check{K}} =  &   \check{I} + \left( e^x - 1  \right) \check{K} +  Q_2(x) S^2 \check{D}^2 + \\
& Q_3(x) S^3\check{D}^3 + Q_4(x)  S^4 \check{D}^4 + Q_5(x)  S^5 \check{D}^5 + \cdots  \\
\end{eqnarray*} 
or
\begin{equation}
e^{ x \check{K}} = \sum_{n=0}^{\infty} Q_n(x) S^n \check{D}^n 
\end{equation}
where $\check{D}^0 = \check{I}$ and
\begin{equation}
Q_2(x) = \frac{1}{2!} x^2  +  \frac{1}{3!} x^3 \alpha_{3, 2} + \frac{1}{4!} x^4 \alpha_{4, 2} +  \frac{1}{5!} x^5    \alpha_{5, 2} +    \cdots,
\end{equation}
\begin{equation}
Q_3(x) =  \frac{1}{3!} x^3  + \frac{1}{4!} x^4 \alpha_{4, 3} +  \frac{1}{5!} x^5    \alpha_{5, 3} +    \cdots 
\end{equation}
\begin{equation}
Q_4(x) = \frac{1}{4!} x^4 + \frac{1}{5!} x^5    \alpha_{5, 4} +    \cdots,
\end{equation}
\begin{equation}
Q_5(x) = \frac{1}{5!} x^5   +  \cdots.
\end{equation}
\begin{equation*} 
\vdots 
\end{equation*}
Now, one can use (\ref{new_exponential}) to compute the exact solution in terms of the Greeks.
 
\section{The exact solution in terms of the Greeks}
Substituting (\ref{new_exponential}) in the exact solution  (\ref{exactsolution}) gives
\begin{equation} \label{exactsolution2}
\pi(S, \tau) =\left\{\begin{array}{ll}
C(S, \tau) & 0 < \tau < \tau_{1} \\
\sum_{n=0}^{\infty} e^{ - v_0 (\tau - \tau_1) }  Q_n(v_0 (\tau - \tau_1)) S^n \check{D}^n \ C(S, \tau)& \tau_{1}< \tau < \tau_{2} \\
\sum_{n=0}^{\infty} e^{ - v_0 (\tau_2 - \tau_1) }  Q_n(v_0 (\tau_2 - \tau_1)) S^n \check{D}^n \ C(S, \tau) & \tau_{2}< \tau < T
\end{array}\right.
\end{equation} \\
Now

\begin{equation}
\begin{array} {ll}
\sum_{n=0}^{\infty} e^{ - v_0 (\tau - \tau_1) }  Q_n(v_0 (\tau - \tau_1)) S^n \check{D}^n \ C(S, \tau) = & e^{ - v_0 (\tau - \tau_1) } C(S, \tau)  +  \\
& e^{ - v_0 (\tau - \tau_1) } \left( e^{v_0 (\tau - \tau_1) }  - 1   \right) S \frac{\partial C(S, \tau)}{\partial S} + \\
&     e^{ - v_0 (\tau - \tau_1) }  Q_2(v_0 (\tau - \tau_1)) S^2  \frac{\partial^2 C(S, \tau)}{\partial S^2} +  \\
&    e^{ - v_0 (\tau - \tau_1) } Q_3(v_0 (\tau - \tau_1))   S^3 \frac{\partial^3 C(S, \tau)}{\partial S^3} + \cdots    
\end{array} 
\end{equation}
that is
\begin{equation}
\begin{array} {ll}
\sum_{n=0}^{\infty} e^{ - v_0 (\tau - \tau_1) }  Q_n(v_0 (\tau - \tau_1)) S^n \check{D}^n \ C(S, \tau) = & e^{ - v_0 (\tau - \tau_1) } C(S, \tau)  +  \\
& \left( 1- e^{- v_0 (\tau - \tau_1) } \right) S \ \Delta(S, \tau) + \\
&     e^{ - v_0 (\tau - \tau_1) }  Q_2(v_0 (\tau - \tau_1)) \ S^2 \  \Gamma(S, \tau) +  \\
&    e^{ - v_0 (\tau - \tau_1) } Q_3(v_0 (\tau - \tau_1)) \ S^3 \ Spd(S, \tau) + \cdots    
\end{array} 
\end{equation}
So finally, the exact solution for the square bubble can be written as
\begin{equation} \label{exactsolution4}
\pi(S, \tau) =\left\{\begin{array}{ll}
C(S, \tau) & 0 < \tau < \tau_{1} \\
   &   \\
e^{ - v_0 (\tau - \tau_1) } C(S, \tau)  + \left( 1- e^{- v_0 (\tau - \tau_1) } \right) S \ \Delta(S, \tau) + & \\
 e^{ - v_0 (\tau - \tau_1) }  Q_2(v_0 (\tau - \tau_1)) \ S^2 \ \Gamma(S, \tau) + &   \\
e^{ - v_0 (\tau - \tau_1) } Q_3(v_0 (\tau - \tau_1)) \ S^3 \ Spd(S, \tau) + \cdots & \tau_{1}< \tau < \tau_{2} \\
  &   \\
e^{ - v_0 (\tau_2 - \tau_1) } C(S, \tau)  + \left( 1- e^{- v_0 (\tau_2 - \tau_1) } \right) S \ \Delta(S, \tau) + & \\
e^{ - v_0 (\tau_2 - \tau_1) }  Q_2(v_0 (\tau_2 - \tau_1)) \ S^2 \ \Gamma(S, \tau) + &   \\
e^{ - v_0 (\tau_2 - \tau_1) } Q_3(v_0 (\tau_2 - \tau_1)) \ S^3 \ Spd(S, \tau) + \cdots & \tau_{2}< \tau < T \\
\end{array}\right.
\end{equation} \\
One can use the above expression to find an approximation for $\pi(S,t)$.
To do this, one can proceed in the following way: as the exponential $e^{ - v_0 (\tau - \tau_1) }$ is easy to compute, one can maintain it without any change (that is, one can consider it to all orders in perturbation theory).
Instead, the the functions $Q_j$ will be truncated at some order.
For example, if one cuts the series in (\ref{exactsolution4}) at order three in $S$, the function $Q_3$ will be truncated at third order too.
The same will be valid for $Q_2$.
Thus
\begin{equation}
Q_3(x) \approx \frac{1}{3!} x^3,
\end{equation}
\begin{equation}
Q_2(x) \approx \frac{1}{2!} x^2  +  \frac{1}{3!} x^3 \alpha_{3, 2}.
\end{equation}
Note that the $Q_j$ for $j > 3$ will not appear  in this approximation, because they all started with powers in $v_0$ of order four and higher.
Thus, we consider the following approximation for the option price in the presence of a square bubble.
\begin{equation} \label{exactsolution5}
\pi(S, \tau) \approx \left\{\begin{array}{ll}
C(S, \tau) & 0 < \tau < \tau_{1} \\
&   \\
e^{ - v_0 (\tau - \tau_1) } C(S, \tau)  + \left( 1- e^{- v_0 (\tau - \tau_1) } \right) S \ \Delta(S, \tau) + & \\
e^{ - v_0 (\tau - \tau_1) } \Big(   \frac{1}{2!} v_0^2 (\tau - \tau_1)^2  +  \frac{1}{3!} v_0^3 (\tau - \tau_1)^3 \alpha_{3, 2} \Big)  \ S^2 \ \Gamma(S, \tau) + &   \\
e^{ - v_0 (\tau - \tau_1) } \Big(  \frac{1}{3!} v_0^3 (\tau - \tau_1)^3   \Big) \ S^3 \ Spd(S, \tau)  & \tau_{1}< \tau < \tau_{2} \\
&   \\
e^{ - v_0 (\tau_2 - \tau_1) } C(S, \tau)  + \left( 1- e^{- v_0 (\tau_2 - \tau_1) } \right) S \ \Delta(S, \tau) + & \\
e^{ - v_0 (\tau_2 - \tau_1) } \Big(   \frac{1}{2!} v_0^2 (\tau_2 - \tau_1)^2  +  \frac{1}{3!} v_0^3 (\tau_2 - \tau_1)^3 \alpha_{3, 2} \Big)  \ S^2 \ \Gamma(S, \tau) + &   \\
e^{ - v_0 (\tau_2 - \tau_1) } \Big(  \frac{1}{3!} v_0^3 (\tau_2 - \tau_1)^3   \Big) \ S^3 \ Spd(S, \tau) & \tau_{2}< \tau < T \\
\end{array}\right.
\end{equation} \\
Figure 2 shows this last approximation for the same values of $f_0/\sigma$ as in Figure 1.
\begin{figure}[h!]
	\centering
	\includegraphics[scale=1.7]{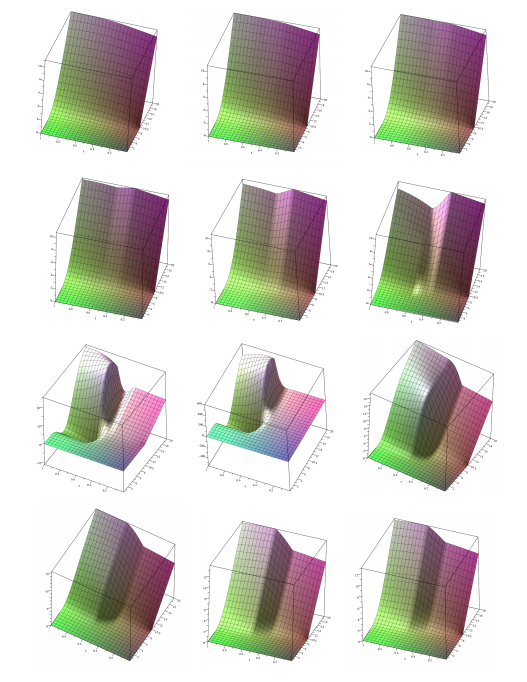}
	\caption{from left to right and from top to the bottom, approximate (\ref{exactsolution5}) Call solution for $\frac{f_0}{\sigma}$ = 0, 0.1, 0.2,  0.4, 0.6, 0.8, 0.9, 0.95, 1.10, 1.20, 1.40, 1.80  respectively.}
\end{figure}
\noindent \\

\section{The higher interaction limit $f \rightarrow \infty$}
In this section the  limit $f \rightarrow \infty$  for the arbitrage bubble will be considered.
Note that from equation (\ref{potentialSt}), 
\begin{equation}
\lim_{f \rightarrow \infty}  v(S,t) = \lim_{f \rightarrow \infty}  \frac{(r-\alpha) f(S, t)}{\sigma-f(S, t)} = -(r-\alpha),
\end{equation}
so the Black--Scholes equation (\ref{BSEinteraction}) in this limit is
\begin{equation} 
\frac{\partial \pi}{\partial t}+\frac{1}{2} \sigma^{2} S^{2} \frac{\partial^{2} \pi}{\partial S^{2}}+r\left(S \frac{\partial \pi}{\partial S}-\pi\right) -(r-\alpha) \left(S \frac{\partial \pi}{\partial S}-\pi\right)=0,
\end{equation}
that is
\begin{equation} 
\frac{\partial \pi}{\partial t}+\frac{1}{2} \sigma^{2} S^{2} \frac{\partial^{2} \pi}{\partial S^{2}}+ \alpha \left(S \frac{\partial \pi}{\partial S}-\pi\right) = 0.
\end{equation}
This is a free Black--Scholes equation with an interest rate equal to $\alpha$ instead of $r$.
Thus, the higher interaction limit of this theory is free theory again.
For the case of a call, Figure 3 shows the free solutions for $r = 0.2$ and $r = \alpha = 0.8$.
Note that both surfaces are smooth.
The approximate solutions for higher values of $f=h$ in Figures 1 or 2 do not recover this smooth behaviour.
This is because it requires the limit $h \rightarrow \infty$, where the approximate solution is not longer valid.
So, how can achieve the limit $h \rightarrow \infty$ using Dirac's picture?
\begin{figure}[h!]
	\centering
	\includegraphics[scale=0.95]{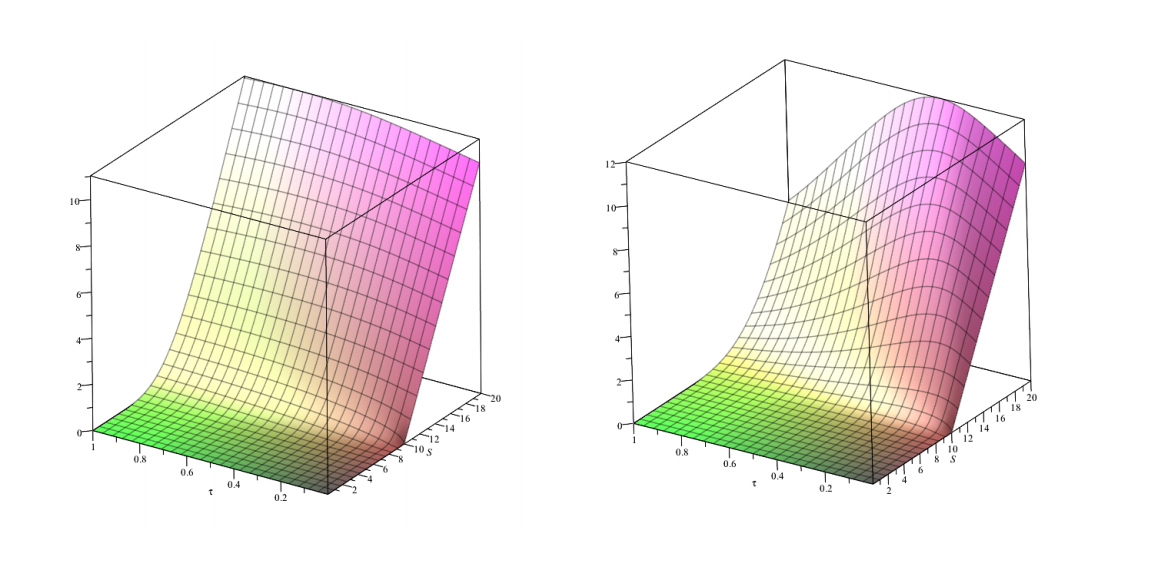}
	\caption{ Free Black--Scholes solutions for a call with $r=0.2$  (left) and $r=\alpha=0.8$ (right).}
\end{figure}
\noindent \\ 
First, one notes that the interacting Black--Scholes (\ref{BSEinteraction}) equation can be written as 
\begin{equation} 
\frac{\partial \pi}{\partial t}+\frac{1}{2} \sigma^{2} S^{2} \frac{\partial^{2} \pi}{\partial S^{2}}+ (\alpha + r - \alpha ) \left(S \frac{\partial \pi}{\partial S}-\pi\right) + v(S,t) \left(S \frac{\partial \pi}{\partial S}-\pi\right)=0,
\end{equation}
or
\begin{equation} 
\frac{\partial \pi}{\partial t}+\frac{1}{2} \sigma^{2} S^{2} \frac{\partial^{2} \pi}{\partial S^{2}}+ \alpha \left(S \frac{\partial \pi}{\partial S}-\pi\right) + \Big( (r - \alpha ) + v(S,t) \Big) \left(S \frac{\partial \pi}{\partial S}-\pi\right)=0,
\end{equation}
by defining the potential $v^*(S,t)$ by
\begin{equation}
v^*(S,t) = (r - \alpha ) + v(S,t) =  (r - \alpha ) + \frac{(r-\alpha) f(S, t)}{\sigma-f(S, t)} = \frac{(r-\alpha) \sigma}{\sigma-f(S, t)} 
\end{equation}
so the Black--Scholes equation becomes 
\begin{equation} 
\frac{\partial \pi}{\partial t}+\frac{1}{2} \sigma^{2} S^{2} \frac{\partial^{2} \pi}{\partial S^{2}}+ \alpha \left(S \frac{\partial \pi}{\partial S}-\pi\right) + v^*(S,t) \left(S \frac{\partial \pi}{\partial S}-\pi\right)=0.
\end{equation}
Note that the potentials $v$ and $v^*$ can be written as 
\begin{equation} v(S,t) = \frac{(r-\alpha) h(S,t) }{1-h(S,t)} 
\end{equation}
\begin{equation} v^*(S,t) = \frac{(\alpha - r) h^*(S,t) }{1-h^*(S,t)} 
\end{equation}
with 
\begin{equation}
h(S,t) = \frac{f(S,t)}{\sigma} = \frac{1}{\frac{\sigma}{f(S,t)}} = \frac{1}{h^*(S,t)}
\end{equation}
where $h(S,t)$ gives the amplitude of the bubble as a fraction of the volatility parameter.
Note that lower values of $h(S,t)$ imply higher values of $h^*(S,t)$ and vice versa.\\ \\
In this way, one can write the interacting Black--Scholes equation explicitly as
\begin{equation} \label{BSEinteractionLOW}
\frac{\partial \pi}{\partial t}+\frac{1}{2} \sigma^{2} S^{2} \frac{\partial^{2} \pi}{\partial S^{2}}+r\left(S \frac{\partial \pi}{\partial S}-\pi\right) + \frac{(r-\alpha) h(S,t) }{1-h(S,t)}  \left(S \frac{\partial \pi}{\partial S}-\pi\right)=0,
\end{equation}
as well as
\begin{equation} \label{BSEinteractionHIGH}
\frac{\partial \pi}{\partial t}+\frac{1}{2} \sigma^{2} S^{2} \frac{\partial^{2} \pi}{\partial S^{2}}+\alpha\left(S \frac{\partial \pi}{\partial S}-\pi\right) + \frac{(\alpha - r) h^*(S,t) }{1-h^*(S,t)}  \left(S \frac{\partial \pi}{\partial S}-\pi\right)=0.
\end{equation}
One can think of equation (\ref{BSEinteractionLOW}) as representing a `low-energy' form of the interacting Black--Scholes  equation and (\ref{BSEinteractionHIGH}) as its `high-energy' form.
Note that (\ref{BSEinteractionLOW}) and (\ref{BSEinteractionHIGH}) are the same equation, so the interacting Black--Scholes equation is invariant under the discrete transformation
\begin{equation} \label{discretetransformation}
r \leftrightarrows \alpha, \ \ \ \ \ \ \  h \leftrightarrows h^*
\end{equation}
and this transformations maps $v$ in $v*$ and vice-versa.
This implies that the set of all solutions of the interacting Black--Scholes equation is preserved under this transformation. Thus, solutions of (\ref{BSEinteractionLOW}) can be obtained from solutions of (\ref{BSEinteractionHIGH}) using transformation (\ref{discretetransformation}) and vice versa.  \\ \\
Also, due to the fact that $ h^* = \frac{1}{h}$, the low coupled limit $ h << 1$ of the `low-energy' form corresponds to the high coupled limit $h^* >> 1$ of the `high-energy' form.
In the same way, the higher coupled limit $h >> 1$ of the `low-energy' form correspond to the low coupled limit $ h^* << 1$ of the `high-energy' form.\\ \\
Thus, one can study the higher interacting limit near $h = \infty$ of the `low-energy' form, by analysing (via Dirac's picture) the low interacting limit $ h^* << 1$ of the `high-energy' form.\\ 
To do that, one can use the invariance of the solutions of B--S under the transformation (\ref{discretetransformation}).
In fact, for the square bubble, the exact solution (\ref{exactsolution4}) is the `low-energy' form.
To find its `high-energy'  form, one perform the transformation (\ref{discretetransformation})  on (\ref{exactsolution4}), that is $v \rightarrow v^*$, to give 
\begin{equation} \label{exactsolution4highform}
 \pi(S, \tau) =\left\{\begin{array}{ll}
 C^*(S, \tau) & 0 < \tau < \tau_{1} \\
 &   \\
 e^{ - v^*_0 (\tau - \tau_1) } C^*(S, \tau)  + \left( 1- e^{- v^*_0 (\tau - \tau_1) } \right) S \ \Delta^*(S, \tau) + & \\
 e^{ - v^*_0 (\tau - \tau_1) }  Q_2(v^*_0 (\tau - \tau_1)) \ S^2 \ \Gamma^*(S, \tau) + &   \\
 e^{ - v^*_0 (\tau - \tau_1) } Q_3(v^*_0 (\tau - \tau_1)) \ S^3 \ Spd^*(S, \tau) + \cdots & \tau_{1}< \tau < \tau_{2} \\
 &   \\
 e^{ - v^*_0 (\tau_2 - \tau_1) } C^*(S, \tau)  + \left( 1- e^{- v^*_0 (\tau_2 - \tau_1) } \right) S \ \Delta^*(S, \tau) + & \\
 e^{ - v^*_0 (\tau_2 - \tau_1) }  Q_2(v^*_0 (\tau_2 - \tau_1)) \ S^2 \ \Gamma^*(S, \tau) + &   \\
 e^{ - v^*_0 (\tau_2 - \tau_1) } Q_3(v^*_0 (\tau_2 - \tau_1)) \ S^3 \ Spd^*(S, \tau) + \cdots & \tau_{2}< \tau < T \\
 \end{array}\right.
 \end{equation} \\
 where one changed 
 \begin{equation}
 v_0 = \frac{(r-\alpha) f_0}{\sigma-f_0} = \frac{(r-\alpha) h_0}{1-h_0},  \ \ \ \ \ h_0 = \frac{f_0}{\sigma}
 \end{equation}
 in (\ref{exactsolution4}) by 
 \begin{equation}
 v^*_0 = \frac{(\alpha - r ) h^*_0}{1 - h^*_0}  \ \ \ \ h^*_0 = \frac{\sigma}{f_0}
 \end{equation}
 and $C^*$, $\Delta^*$, $\Gamma^*$, $Spd^* \cdots$ are the same $C$, $\Delta$, $\Gamma$, $Spd \cdots$ in which one replaces $r$ by $\alpha$.
By keeping terms up $S^3$ in (\ref{exactsolution4highform}) one has the approximate solution  
 \begin{equation} \label{exactsolution5highform}
 \pi(S, \tau) \approx \left\{\begin{array}{ll}
 C^*(S, \tau) & 0 < \tau < \tau_{1} \\
 &   \\
 e^{ - v^*_0 (\tau - \tau_1) } C^*(S, \tau)  + \left( 1- e^{- v^*_0 (\tau - \tau_1) } \right) S \ \Delta^*(S, \tau) + & \\
 e^{ - v^*_0 (\tau - \tau_1) } \Big(   \frac{1}{2!} (v^*_0)^2 (\tau - \tau_1)^2  +  \frac{1}{3!} (v^*_0)^3 (\tau - \tau_1)^3 \alpha_{3, 2} \Big)  \ S^2 \ \Gamma^*(S, \tau) + &   \\
 e^{ - v^*_0 (\tau - \tau_1) } \Big(  \frac{1}{3!} (v^*_0)^3 (\tau - \tau_1)^3   \Big) \ S^3 \ Spd^*(S, \tau)  & \tau_{1}< \tau < \tau_{2} \\
 &   \\
 e^{ - v^*_0 (\tau_2 - \tau_1) } C^*(S, \tau)  + \left( 1- e^{- v^*_0 (\tau_2 - \tau_1) } \right) S \ \Delta^*(S, \tau) + & \\
 e^{ - v^*_0 (\tau_2 - \tau_1) } \Big(   \frac{1}{2!} (v^*_0)^2 (\tau_2 - \tau_1)^2  +  \frac{1}{3!} (v_0)^3 (\tau_2 - \tau_1)^3 \alpha_{3, 2} \Big)  \ S^2 \ \Gamma^*(S, \tau) + &   \\
 e^{ - v^*_0 (\tau_2 - \tau_1) } \Big(  \frac{1}{3!} (v^*_0)^3 (\tau_2 - \tau_1)^3   \Big) \ S^3 \ Spd^*(S, \tau) & \tau_{2}< \tau < T \\
 \end{array}\right.
 \end{equation} \\
 which is valid in the the high interaction limit $f_0 \rightarrow \infty,  \ \ v_0 \rightarrow \infty $.
Figure 4 shows the approximate solution for several values of $h^*$.
\begin{figure}[h!]
	\centering
	\includegraphics[scale=1.5]{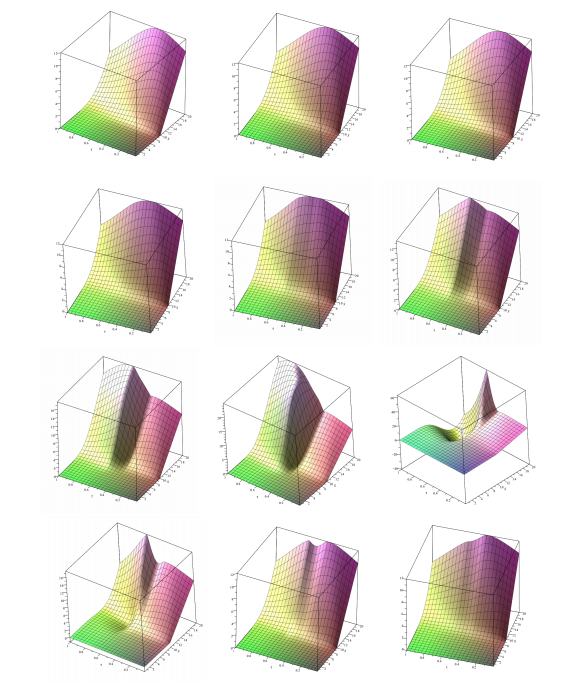}
	\caption{from left to right and from top to the bottom, the approximate high energy form (\ref{exactsolution5highform}) of the call solution for $h^*$ = 0, 0.1, 0.2,  0.4, 0.6, 0.8, 0.9, 0.95, 1.10, 1.20, 1.40, 1.80  respectively.}
\end{figure}
\newpage
\section{Conclusions}
In this article, Dirac's interaction picture of quantum mechanics is applied to obtain an approximate solution of the interacting Black--Scholes equation in the presence of arbitrage bubbles.
For a square bubble, an initial naive approximate solution is presented in terms of the first three Greeks.\\ \\
After that, an exact solution is found in terms of all high-order Greeks.
By truncating this solution, a more exact approximation is obtained that includes some terms at all orders in perturbation theory.\\ \\
Also, some properties of the interacting Black--Scholes are analysed.
It is found that this model is invariant under a discrete transformation that interchanges the interest rate with the mean of the asset price.
This implies that the interacting Black--Scholes equation can be written in two different ways: low energy and high energy forms.\\ \\
For the low energy form, the analysis is done by perturbing the free solution of with the value of the interest rate being $r$, which is valid for $v_0 = \frac{(r -\alpha) h}{(1-h)} << 1$ where $h = \frac{f_0}{\sigma} << 1$ and $f_0$ is the height of the square bubble.\\
For the opposite case $h=\frac{f_0}{\sigma} >> 1$, that is, the high interaction limit, one can use the dual high energy description in terms of the parameter  $v^*_0 = \frac{(\alpha - r) h^*}{(1-h^*)} << 1$ with $h^*=\frac{1}{h} << 1$ by perturbing the free B--S solution for the interest rate $\alpha$.\\ \\
In a later article, I will provide a detailed study and comparison of the approximate solution (\ref{exactsolution5}) and the exact numerical solution of (\ref{BSEinteractionv3}) to establish a relation between the error given by the difference of both solutions and the order of the truncated perturbation series in some order $n$.

\end{document}